\documentclass[aps,twocolumn,showpacs,amsmath,amsfonts,amssymb]{revtex4-2}
\usepackage{graphicx}
\usepackage{epsf}
\usepackage{epstopdf}
\begin{document}

\title{ Exact Solutions for Solitary Waves in a Bose-Einstein Condensate under the Action of a Four-color Optical Lattice
 }

\author{Barun Halder$^{\mathrm{1}}$\footnote{...............}}
\author{Suranjana Ghosh$^{\mathrm{2}}$\footnote{...............}}
\author{Pradosh Basu$^{\mathrm{1}}$\footnote{................}}
\author{Jayanta Bera$^{\mathrm{1}}$\footnote{................}}
\author{Boris Malomed$^{\mathrm{3,4}}$\footnote{malomed@tauex.tau.ac.il}}
\author{Utpal Roy$^{\mathrm{1}}$\footnote{uroy@iitp.ac.in}}

\affiliation{$^{\mathrm{1}}$Indian Institute of Technology Patna,
Bihta, Patna-801103, India\\$^{\mathrm{2}}$Indian Institute of
Science Education and Research Kolkata, Mohanpur-741246, India\\$^{\mathrm{3}}$Department of Physical Electronics, School of Electrical Engineering, Faculty of Engineering, and Center for Light-Matter Interaction, Tel Aviv University, P.O.B. 39040, Ramat Aviv, Tel Aviv,
Israel\\$^{\mathrm{4}}$Instituto de Alta Investigaci\'{o}n, Universidad de Tarapac\'{a}, Casilla 7D, Arica, Chile}

\begin{abstract}
We address dynamics of Bose-Einstein condensates (BECs) loaded into a one-dimensional four-color optical lattice (FOL) potential with commensurate wavelengths and tunable intensities. The analysis identifies specific multi-parameters forms of the FOL potential which admits exact solitary-wave solutions. This newly found class of potentials includes more
particular species, such as frustrated double-well superlattices and bichromatic and three-color lattices. Our exact solutions provide options for controllable positioning of density maxima of the localized patterns, and tunable Anderson-like localization in the frustrated potential. A numerical analysis is performed to establish dynamical stability and structural stability of the obtained solutions, which makes them relevant for experimental realization. The newly found solutions offer applications to the design of schemes for quantum simulations and processing quantum information.
\end{abstract}


\pacs{.................} \maketitle
\section{Introduction}
A suitably prepared standing wave of laser radiation can form an optical
lattice (OL), which are broadly used for trapping and steering of ultracold
atoms \cite{Bloch,Greiner,Denschlag,Jaksch,Konotop,Oberthaler,Lewenstein,Dutta}.
Offering a versatile platform for research in the area of matter waves, OLs
have become the most appropriate candidate for the realization of quantum
simulations \cite{simulator,Gross,Schafer}. Further, ultracold atoms and
Bose-Einstein condensates (BECs) trapped in an OL are used as a basis for
the development of atomic clocks, quantum sensors, quantum computers, and a
variety of other applications in quantum technologies \cite%
{Brennen,Katori,Wang}.

In particular, the study of BEC under the action of geometrically frustrated
OLs has drawn much interest \cite{Schulte,Adhikari,Nat,Yamamoto}. Many
complex phenomena have been found in this connection, including
Anderson-like localization and negative absolute temperature \cite%
{Nat,Billy,Braun,Nath}. Optical superlattices subjected to frustration offer
a potential for the development of tools which can hold and mould robust
matter-wave states, such as solitons \cite{Cheng,Das,EPJD,Sun,Li}.
Theoretical studies in this direction are chiefly limited to a variety of
bi-color optical lattices (BOL). A more general form of multi-color OLs may
offer additional advantages, including the following points: (i) the color
(wavelength) and intensity of the constituent beams, building the effective
optical potential, greatly influence the manner in which the atoms are
trapped; (ii) the formation of solitons requires a specific correlation
between the nonlinearity and the trap parameters, which the multi-color OL
may help to maintain; (iii) relations between intensities of the constituent
beams may be used to optimize the creation of the self-trapped patterns.
Thus, multi-color beams can be used to design potentials necessary for
holding complex soliton patterns.

The aim of this work is to introduce a four-color OL (FOL) with commensurate
wavelengths, which acts on a cigar-shaped (quasi-one-dimensional) BEC with
the cubic nonlinearity. The corresponding Gross-Pitaevskii equation (GPE) is
used to find appropriate relations between the nonlinearity and the
potential parameters which help to support solitons. We produce analytical
solutions which identify the specific form of the FOL and its parameter
domain which provide tunability of the soliton-building scheme. Many exact
condensate wave functions are obtained, and the results are illustrated by
several characteristic examples. These solutions may be used for
applications similar to those proposed in previous works \cite{Malomed,Qiu,Wind,Ghosh,Yang,PRA}. Stability of the exact wave functions is
addressed by means of direct simulations, adding random perturbations either
to the underlying stationary solution, or to the external trap (the latter
implies the consideration of the structural stability of the exact
solutions). We thus find that our solutions are fully stable, both
dynamically and structurally.

\section{Exact Analytical Model for Obtaining the Solitary Excitations under
the Novel FOL trap}
The FOL potential is produced by the combinations of four OLs with
commensurate wave numbers, $l$, $2l$, $3l$, and $4l$, while the
corresponding intensities of the laser beams, $V_{1,2,3,4}$, are treated as
free parameters, with the intention to find appropriate relations between
them. The corresponding effective potential acting on atoms is
\begin{equation}\label{Gpot}
\hspace{1cm}V(z)=\sum_{j=1}^{4}V_{j}\cos (jlz).
\end{equation}
The lattice depth may be compared to the recoil energy, $E_{R}=2\pi
^{2}\hbar ^{2}/\left( M\lambda ^{2}\right) ,$ and the scaled lattice
wave-vector is given by $l=2\pi a_{\perp}/\lambda $, where $\lambda $ is the
wavelength, $M$ is the mass of the BEC atom, $a_{\perp}=(\hbar/\left(M\omega_{\perp}\right))^{1/2}$ and $\omega_{\perp}$ is the transverse frequency . The dimensionless 1D-GPE has the form
\begin{equation}
\hspace{-0.5cm}\left[ i\frac{\partial }{\partial t}+\frac{1}{2}\frac{%
\partial ^{2}}{\partial z^{2}}-g(z,t)|\psi (z,t)|^{2}-V(z)-i{\tau }(z,t)%
\right] \psi (z,t)=0. \label{GP}
\end{equation}%
Here, $g(z,t)$ is the nonlinearity coefficient, which may be made space- and
time-modulated, while $\tau (z,t)$ represents the space- and time-modulated
loss or gain of the condensate atoms. For illustration, we have exploited
experimentally feasible parameters of Li$^{7}$ BEC in the quasi-1D trapping
configuration with transverse frequency $\omega _{\perp }=2\pi \times 710$
Hz, OL wavelength $\lambda =10.62$ $\mu $m, and scattering length $%
a_{s}=-0.21$ nm corresponding to attractive interactions between atoms \cite%
{Khaykovich}. By varying the applied magnetic field and angle between the
overlapping laser beams, it is possible to engineer the shape of the
external potential \cite{Inouye,Roberts}.

\begin{figure*}[tbp]
\centering
\includegraphics[width=3. in, scale=.6]{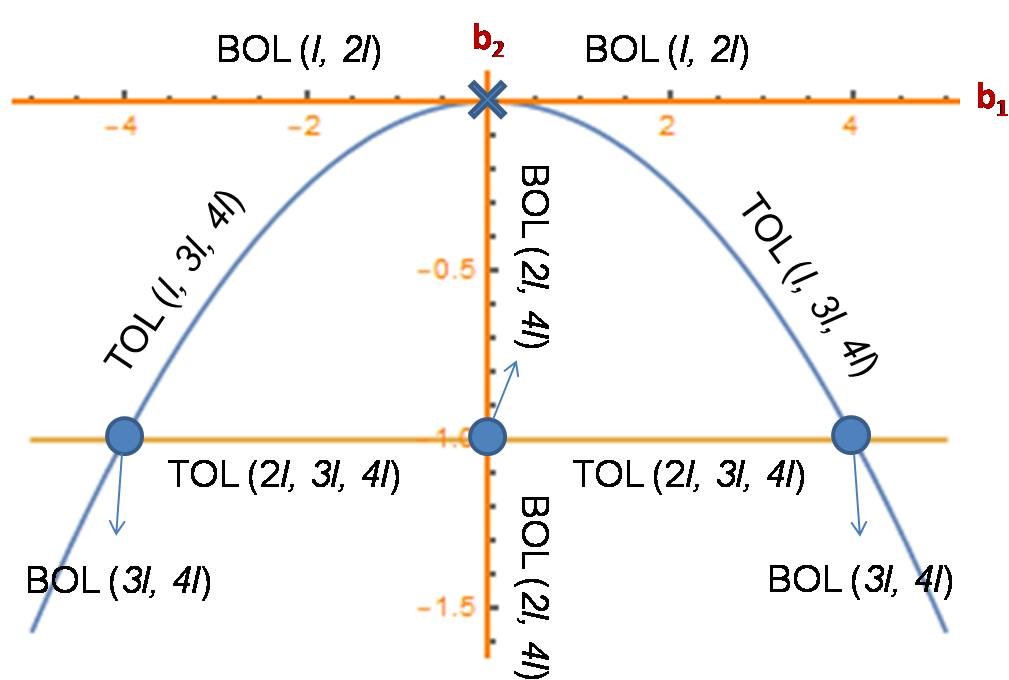}
\caption{Curves and points for $l=0.84$ where the potential is not a FOL,
but a TOL or a BOL. `$\times $' signifies no potential for $b_{1}=b_{2}=0$.
All other points in the $\left( b_{1},b_{2}\right) $ plane correspond to
FOLs.}
\label{b1b2}
\end{figure*}

To produce a spatially localized solution of Eq.~(\ref{GP}), following the
general scheme used for engineering matter-wave configurations \cite{Kengne,Ajay}
we choose an ansatz,
\begin{equation}
\hspace{1.2cm}\psi (z,t)=A(z,t)F\left( B(z,t)\right) e^{i\theta (z,t)},
\label{psi}
\end{equation}%
such that the external potential is supposed to be precisely found by exactly solving $B(z,t)$, amplitude $A(z,t)$, phase $\theta (z,t)$ and the condensate form factor $F[B(z,t)]$. We substitude this ansatz into the GPE (\ref{GP}) and separate out the real and imaginary parts. To establish relations between the physically relevant quantities like nonlinearity, amplitude, phase and external trap for a solitary wave solution, the real part can be mapped to the following nonlinear differential equation,
\begin{equation}
\hspace{1.2cm}\frac{\partial ^2F[B(z,t)]}{\partial B(z,t)^2}-GF^3[B(z,t)]=0,
\label{elliptic}
\end{equation}%
which introduces a constant $G=2g(z,t)A^{2}(z,t)/B_{z}^{2}(z,t)$ in the case of
solitary wave excitations. $G$ is $-1$ for attractive and $1$ for repulsive
inter-atomic interactions. The last consistency condition in Eq.~(\ref{elliptic}) is nothing but the elliptic equation, whose exact solutions are well-known in the form of $12$ Jacobian elliptic functions ($cn[z, m]$, $sn[z, m]$ etc), where $m$ is the modulus parameter with $0 \leq m \leq 1$ \cite{Abramowitz}. One can choose various shapes of the elliptic functions from periodic ($m=0$) to localized ($m=1$), by varying the value of its modulus parameter. Here, we will emphasize only on the localized forms of the elliptic functions, \textit{i.e.,} $cn[z,1]=sech[z]$, for the bright soliton case with attractive nonlinearity and $sn[z,1]=tanh[z]$ for the dark soliton case with repulsive nonlinearity. In addition to solving the above mention equation, we also obtain the following consistency relations:
\begin{eqnarray}
GB_{z}^{2}(z,t)-2A^{2}(z,t)g(z,t)=0, &&  \nonumber \\
B_{t}(z,t)+B_{z}(z,t)\theta _{z}(z,t)=0,[A^{2}(z,t)B_{z}(z,t)]_{z}=0 &&  \nonumber
\\
2A(z,t)A_{t}(z,t)+[A^{2}(z,t)\theta _{z}(z,t)]_{z} &&  \nonumber \\
-2\tau (z,t)A^{2}(z,t)=0\; &&  \nonumber \\
\frac{A_{zz}(z,t)}{2A(z,t)}-\frac{\theta _{z}^{2}(z,t)}{2}-\theta
_{t}(z,t)-V(z)=0, &&  \label{consistencies}
\end{eqnarray}%
where the subscripts stand, as usual, for partial derivatives. The above set of equations is solved simultaneously to produce
\begin{eqnarray}
B(z,t)=\frac{c(t)}{A^{2}(z,t)},\;\; &&\theta _{z}(z,t)=-\frac{A_{t}(z,t)}{%
A_{z}(z,t)},  \nonumber \\
g(z,t)=GB_{z}^{2}(z,t)/2A^{2}(z,t),  \label{B_theta_g}
\end{eqnarray}%
where $c(t)$ is an arbitrary positive definite function of time. These
equations indicate a direct dependence of phase and nonlinearity on the
amplitude of the system which will be determined by the trapping potential
through the last equation of system~(\ref{consistencies}).

\begin{figure*}[tbp]
\centering
\includegraphics[width=5. in]{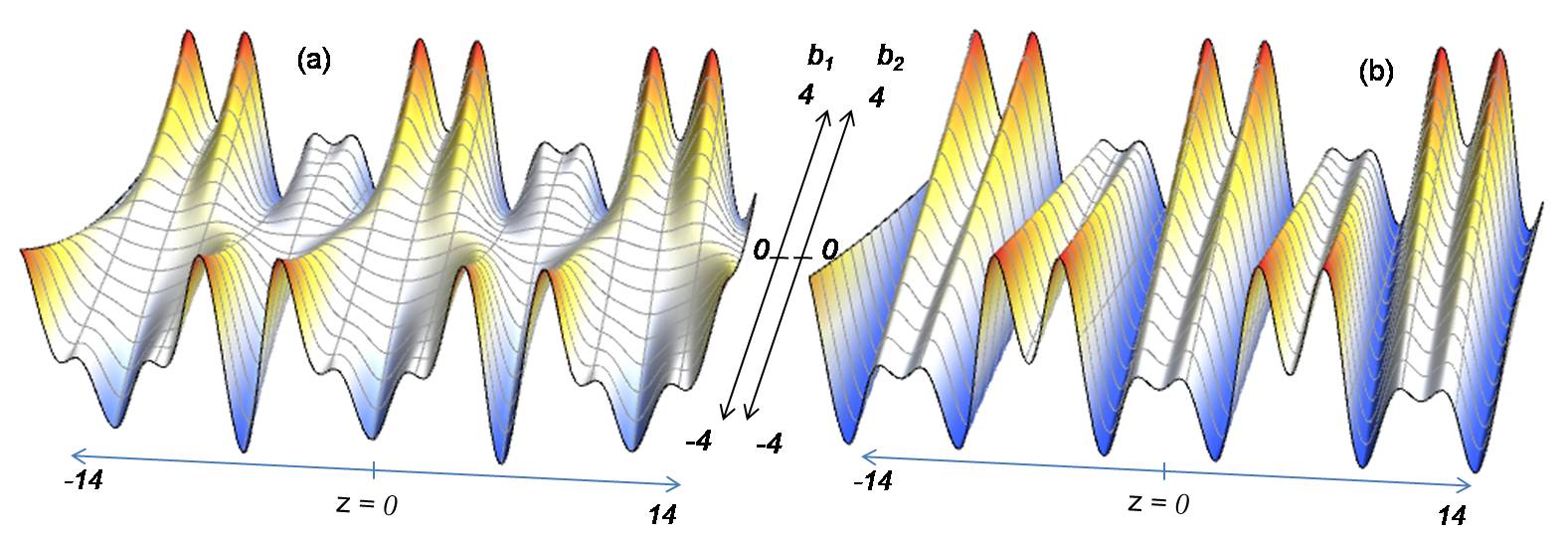}
\caption{The variation of the FOL potential following the change of the
tuning parameters for $l=0.84$: (a) for fixed $b_{2}=2$, $b_{1}$ varies from
$-4$ to $+4$; (b) for fixed $b_{1}=2$, $b_{2}$ varies from$-4$ to $+4$. Here
and in the figures following below, the results are displayed in interval $%
-14<z<+14$.}
\label{pot1}
\end{figure*}

\begin{figure*}[h]
\centering
\includegraphics[width=6. in]{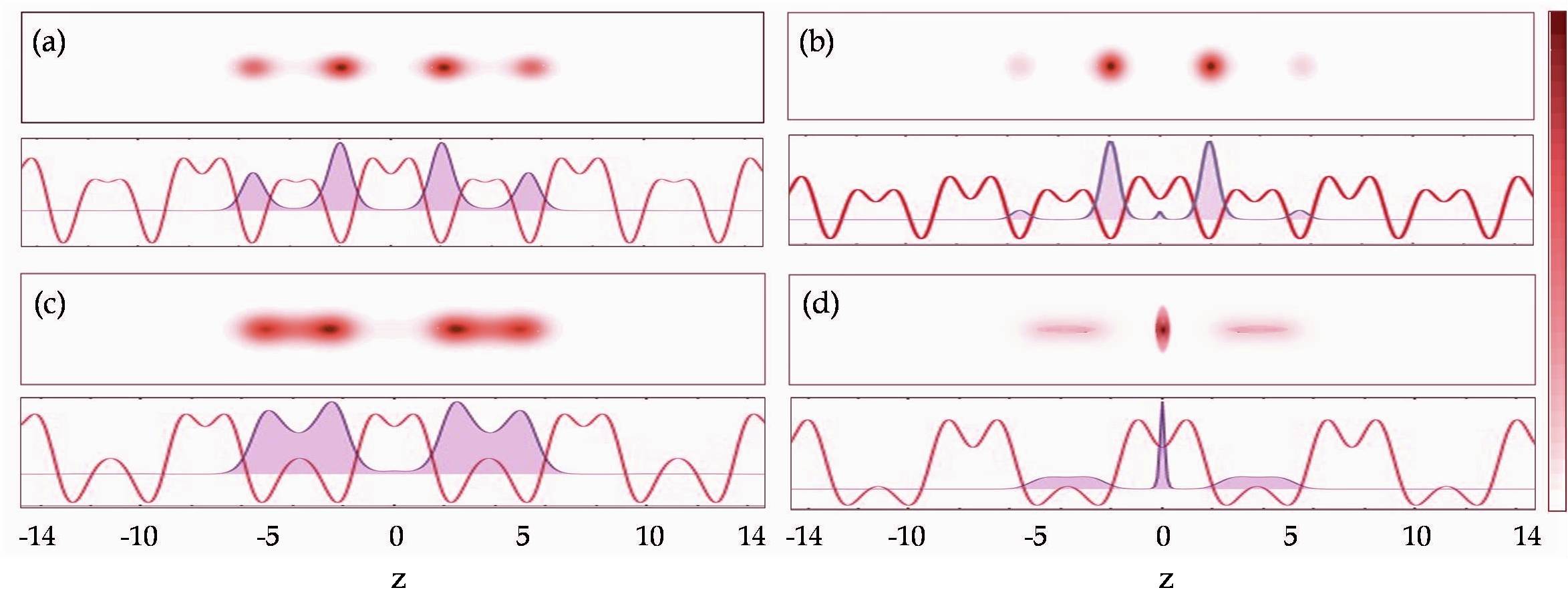}
\caption{Condensate density patterns for $b_{1}>0$ and $b_{2}>0$: (a) $%
b_{1}=1$; $b_{2}=2$, (b) $b_{1}=1$; $b_{2}=3.5$, (c) $b_{1}=2$; $b_{2}=1$,
and (d) $b_{1}=3.5$; $b_{2}=1$. Each plot of (a)-(d) has two panels: the upper panel shows the contour plot of the density and the lower panel consists of a $2$D plot of the density combined with the corresponding potential profile.}
\label{density-positive}
\end{figure*}
\begin{figure*}[h]
\centering
\includegraphics[width=6. in]{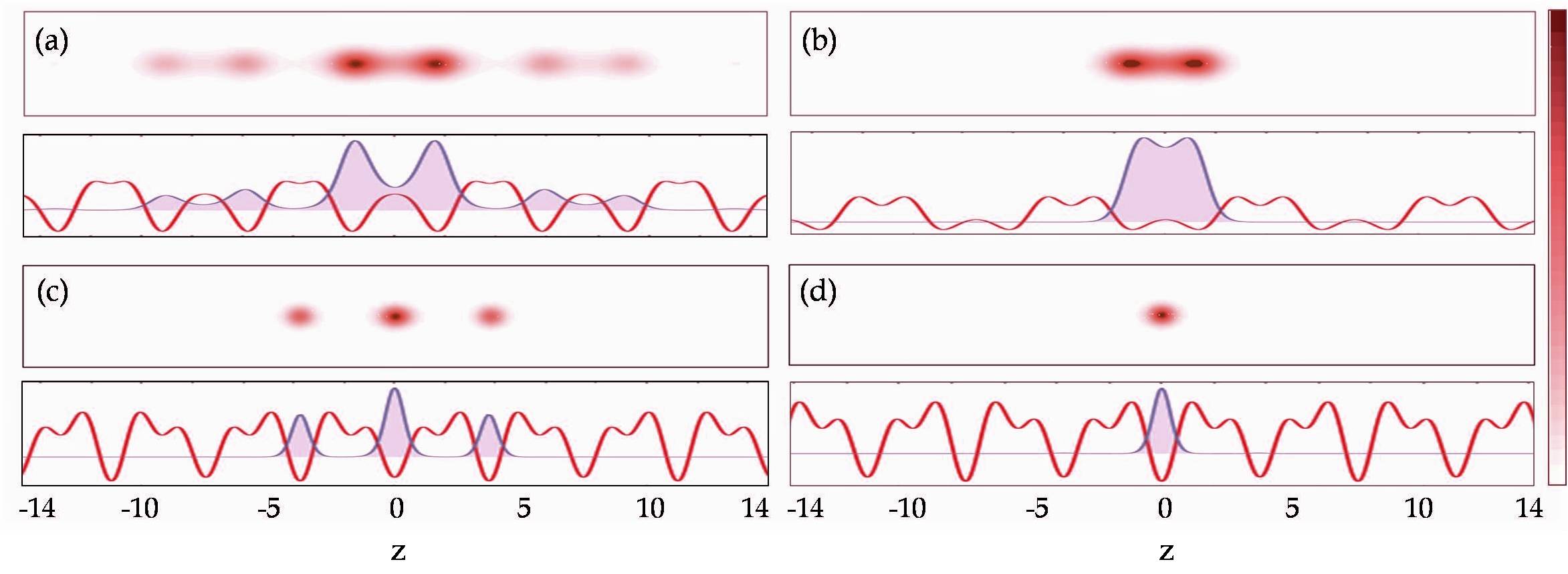}
\caption{Condensate density patterns for $b_{1}<0$ or $b_{2}<0$ or both:
(a) $b_{1}=-1$; $b_{2}=1$, (b) $b_{1}=-3$; $b_{2}=1$, (c) $b_{1}=1$; $%
b_{2}=-3$, and (d) $b_{1}=-1$; $b_{2}=-3$. Each plot of (a)-(d) has two panels: the upper panel shows the contour plot of the density and the lower panel consists of a $2$D plot of the density combined with the corresponding potential profile.}
\label{density-negative}
\end{figure*}

We substitute the expression of the external potential from Eq.~(\ref{elliptic}) into the set of equations (\ref{consistencies}-\ref{B_theta_g}) and obtain the amplitude, phase and nonlinearity in the following exact forms:
\begin{eqnarray}
A(z,t) &=&\sqrt{\frac{c(t)}{\gamma \exp (b_{1}\cos (lz)+b_{2}\cos (2lz))}},
\nonumber \\
\theta (z,t) &=&\frac{1}{16}(l^{2}b_{1}^{2}+l^{2}b_{2}^{2})t,\;\tau (z,t)=%
\frac{1}{2}\frac{c^{\prime }(t)}{c(t)},  \nonumber \\
g(z,t) &=&\frac{G\gamma ^{4}}{2c^{2}(t)}\exp \left( 4b_{1}\cos
(lz)+4b_{2}\cos (2lz)\right) .  \label{A_theta_g}
\end{eqnarray}%
We have here introduced two real constants, $%
b_{1}$ and $b_{2}$, which help us to define the final form of the
FOL-amplitudes:
\begin{eqnarray}
V_{1}=(1+b_{2})\frac{l^{2}b_{1}}{4},\; &&\;V_{2}=\left( \frac{-b_{1}^{2}}{16}%
+b_{2}\right) l^{2},  \nonumber  \label{pot} \\
V_{3}=-\frac{l^{2}b_{1}b_{2}}{4},\; &&\;V_{4}=-\frac{l^{2}b_{2}^{2}}{4}.
\end{eqnarray}%
This is one of the essential results of the present work. Constants $b_{1}$ and $%
b_{2}$ are thus identified as the prime tuning parameters for controlling
the trapping potential and condensate density. For the attractive and
repulsive interactions, assuming the commonly known bright- or dark-soliton
solution of the elliptic equation (Eq.~(\ref{elliptic})), the condensate
wave functions take, severally, the following form:
\begin{widetext}
\begin{eqnarray}
&& \psi (z,t)=\sqrt{\frac{c(t)}{\gamma \exp
\left[ b_{1}\cos (lz)+b_{2}\cos \left( 2lz\right) \right] }}
 sech\Bigg[\gamma \int_{0}^{z}\exp(b_{1}\cos (lz)+b_{2}\cos (2lz))dz^{\prime }\Bigg]\exp (i\theta (z,t)),
\nonumber \\
&& \psi (z,t)=\sqrt{\frac{c(t)}{\gamma \exp (b_{1}\cos
(lz)+b_{2}\cos (2lz))}} \tanh \Bigg[\gamma \int_{0}^{z}\exp (b_{1}\cos
(lz)+b_{2}\cos (2lz))dz^{\prime }\Bigg]\exp (i\theta (z,t))
\end{eqnarray}%
\end{widetext}
We are now in a position to analyze the relevant potential profiles and the
corresponding condensate densities. Condensate densities will be delineated in Fig.~\ref{density-positive} and Fig.~\ref{density-negative} for some parameter domains of $b_1$ and $b_2$ with $c=0.1$, $g=0.1$, and $l=0.84$.

\section{The parameter domain and shape of the tunable FOL}

Figure \ref{b1b2} depicts the structure in the $\left( b_{1},b_{2}\right) $
space, produced by Eq.~(\ref{pot}), where one obtains, as particular cases,
a tri-color optical lattice (TOL), or a BOL. On the contrary, FOL is
obtained in the entire space, excluding the curves and points indicated in
the figure.

The respective FOL potential, drawn in Fig.~\ref{pot1}, seems interesting
enough. For $b_{1}>0$ and $b_{2}>0$, the FOL is a disordered double-well
superlattice, featuring frustrations in terms of both inter- and intra-well separations. Figures~\ref{pot1}(a) and (b) reduce to a BOL at $b_{1}=b_{2}=0$%
. However, the transition to the domain of $b_{1}<0$ or $b_{2}<0$ makes the
potential shapes quite different. In the former case, triple-well superlattice gradually
appears at $b_{1}<0$, whereas in the latter case, a translational shift of the double-wells in the superlattice by half a period is observed. The presently engineered FOL may be the most advanced trapping potential for BEC, derived as an ingredient of exact solutions. It
may find applications to the design of quantum simulation, information and
computation schemes \cite{Qiu,Wind,Ghosh,Yang}. We will further illustrate the results by displaying density patterns.

\section{Density patterns supported by the engineered FOL}

The density patterns in the domain of $b_{1,2}>0$ are displayed in Fig.~\ref%
{density-positive}, along with the respective trapping profile, which help
to understand the formation mechanism for the patterns. The presence of the
inter- and intra-well potential frustration helps one to realize
well-distinguished quantum clouds, that may be employed for the design of
enhanced atom-interferometry (Fig.~\ref{density-positive}(a-c)). When the
intra-well frustration disappears, the previously separated clouds inside
the double well become indistinguishable and the condensate starts
accumulating at the central frustrated site, causing Anderson-like
localization (Fig.~\ref{density-positive}(d)). The wide tunability of the
FOL and the corresponding mesoscopic clouds make it possible to predict a
variety of quantum states, that may be useful for quantum technology \cite%
{Howards,tetrachotomous,Shukla}.

\begin{figure*}[h]
\centering
\includegraphics[width=6. in]{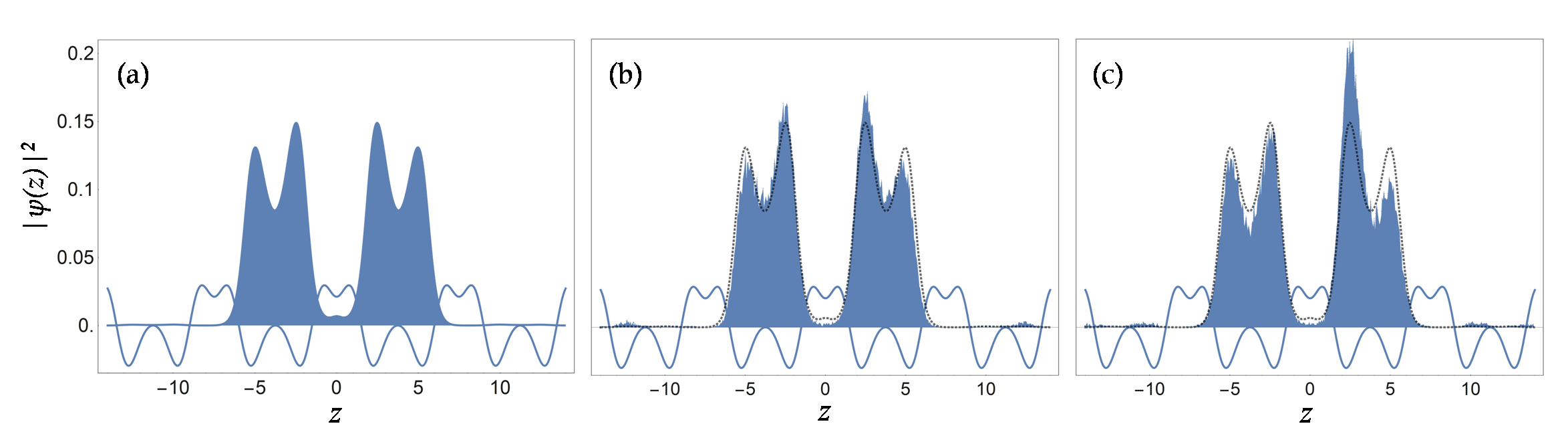}
\caption{Condensate densities are depicted by filled plots at times (a) $t=0$, (b) $t=10ms$, and (c) $t=20ms$, along with the potential energy profile (solid-line curve) for $b_{1}=2$ and $b_{2}=1$. Initial density (dotted curve) is merged with the densities in (b-c) for reference.}
\label{density-time}
\end{figure*}

In Fig.~\ref{density-negative}, we illustrate the situation in the negative
domain: $b_{1}<0$ in Fig.~\ref{density-negative}(a-b), $b_{2}<0$ in Fig.~\ref%
{density-negative}(c), and $b_{1,2}<0$ in Fig.~\ref{density-negative}%
(d). It produces several aligned well-separated spatial Schr\"{o}dinger-cat
states for $b_{1}<0$ \cite{Cirac}. More negative $b_{1}$ offers localization of the
cat-state at the central double well. For $b_{2}<0$, the resulting
triple-well super-lattice generates an odd number of well-separated clouds.
Interestingly, changing the sign of $b_{1}$ at $b_{2}<0$ spatially
translates the triple-well lattice by one period to create a single
BEC-cloud at the center (Fig.~\ref{density-negative}(d)). Thus, a transition
from Fig.~\ref{density-negative}(d) to Fig.~\ref{density-negative}(c) splits
the single cloud into a set of three ones. In addition, a transition from
Fig.~\ref{density-negative}(b) to Fig.~\ref{density-negative}(a) splits the
Schr\"{o}dinger-cat state from one to three. Along with the above-mentioned
possibilities, this scheme of potential engineering offers an efficient
scheme for designing quantum logic gates \cite{Yang,PRA,Zeng,Foot,Vo,Gajdacz}. To illustrate the temporal dynamics of one of the obtained solutions, we choose the trap corresponding to $b_{1}=2$ and $b_{2}=1$ which shows a frustrated double-well super-lattice. Condensate, trapped in this potential, is allowed to evolve in time with a random noise of amplitude $10\%$ of the maximum density.
Condensate densities are depicted in Fig.~\ref{density-time}(a-c) for $t=0$, $t=10ms$, and $t=20ms$, respectively. One can observe that the condensate is maintaining its shape after $t=10ms$, but getting distorted at $t=20ms$.

\begin{figure*}[tbp]
\centering
\includegraphics[width=6. in]{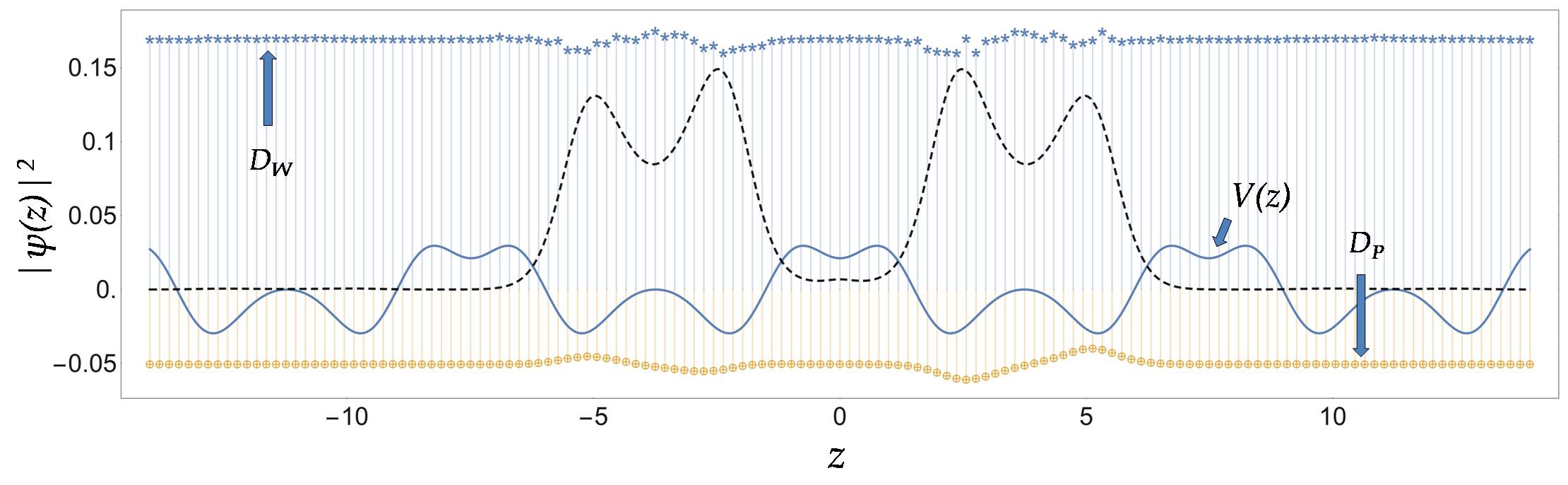}
\caption{The numerical stability analysis of one of the obtained solutions
with $b_{1}=2$ and $b_{2}=1$. The condensate density is depicted by the dotted line and the trap profile, $V(z)$ (not in scale), is superimposed on it
(the solid line). The deviation of the noisy data from their noise-free
counterparts is shown for both kinds of the analyses: the dynamical
stability, $D_{W}$ (the upper curve, composed of symbols $\ast $),
and the structural stability, $D_{P}$ (the lower curve, composed of
symbols $\bigoplus $).}
\label{stability}
\end{figure*}

\section{Dynamical Stability and Structural Stability of the Condensate}
It is obviously necessary to check the dynamical and structural stability of
the special analytical solution produced above. The dynamical stability
pertains to disturbance added to the wave function, while the structural
stability implies deformation of the external trap. We addressed these
problems separately by numerically solving the GPE with the help of the
split-step Fourier method \cite{Nath,EPJD,PRA}. The results are presented in
Fig.~\ref{stability}. In the former case, we have added random white noise $%
\Re _{w}$ to the analytically obtained wave function, while in the latter
case, the noise is added to the external trap. The noisy form of the initial
wave function and potential are represented as
\begin{eqnarray}
\psi _{\mathrm{noisy}}(z,t &=&0)=\psi (z,t=0)+\Re _{w}  \nonumber
\label{noisy} \\
V_{\mathrm{noisy}}(z) &=&V(z)+\Re _{w}.
\end{eqnarray}%
While the stability analysis was performed for a broad range of the
parameters, here we choose $b_{1}=2$ and $b_{2}=1$ for the purpose of
illustration. In Fig.~\ref{stability}, the condensate density profile, along
with the external trap (not in scale), are depicted without the noise. The
wave function is numerically evolved for both the noisy configurations
defined as per Eq.~\ref{noisy}. Amplitude of noise $\Re _{w}$ varies from $0$
to $5\%$ of the maximum amplitude of the initial wave function. In the first scenario, we monitored the evolution of the wave functions, induced by the inputs $\psi _{\mathrm{noisy}}(z,t=0)$ and $\psi(z,t=0)$, with our model potential, $V(z)$. To observe the stability of the stationary state, we computed deviation of thee evolving condensate density ($D_{W}$). In the latter case, we monitored the evolution of the input wave function ($\psi (z,t=0)$) under the action of the potentials $V_{\mathrm{noisy}}(z)$ and $V(z)$, to observe the structural deformation in the condensate density ($D_{P}$). We have simulated the evolution for $10000$ time iterations with properly chosen
space and time steps, $dz=0.277$ $\mu $m and $dt=0.22$ $\mu$s, respectively. The deviation (maximum relative error) of the evolved
noisy data from their noise-free counterparts is shown for both kinds of the
stability analyses in Fig.~\ref{stability} by the upper curve ($\ast $) and
lower one, $\bigoplus $. Observing the noisy density profile after $10000$ iterations, we conclude that the density patterns retain their shapes with minimal deformation, which implies that the analytical solutions are indeed stable against both kinds of the random perturbations (Fig.~\ref{stability}). The observed relative perturbation in the final configurations is near to $1\%$ when the noise is initially added to the wave function, and near to $2\%$ when it is added to the trapping potential. Thus, the presented model and its analytical solutions are physically relevant ones.

\section{Conclusion}

We have reported the exact form of the FOL (four-color optical lattice) trap
for the BEC in one dimension, which makes it possible to produce exact
solutions for the trapped condensate. A variety of experimentally relevant
trap profiles are reported, including one-, two-, three- and four-color OLs
with tunable shapes. It is worthy to stress that there are only two
FOL-tuning parameters, $b_{1}$ and $b_{2}$, instead of four, making the
detailed analysis of the exact solutions feasible. For a chosen trap parameters, the exact condensate density is illustrated and its variations after evolving in time are also shown. By means of systematic simulations, we have established dynamical and structural stability of the exact solutions. The stability against structural perturbations is especially important, as the solutions are valid only for the specially designed form of the FOL potentials. This class of FOL\ trapping potentials offers a straightforward potential for the use in applications, such as quantum simulation and other quantum technologies \cite{Katori,Qiu,Wind,Ghosh,Yang,Howards,tetrachotomous,Shukla}.

  \end{document}